\documentclass[preprint,showpacs,preprintnumbers,
amsmath,showkeys,amssymb,aps,floatfix]{revtex4}

\usepackage{graphicx}
\usepackage{dcolumn}
\usepackage{bm}
\usepackage{longtable}
\usepackage{epsfig}
\usepackage{times}
\usepackage[dvips]{color}

\begin{document}

\title{Electric dipole-forbidden nuclear transitions driven by super-intense laser fields}

\author{Adriana \surname{P\'alffy}}
\email{palffy@mpi-hd.mpg.de}

\author{J\"org \surname{Evers}}
\email{joerg.evers@mpi-hd.mpg.de}

\author{Christoph~H. \surname{Keitel}}
\email{keitel@mpi-hd.mpg.de}

\affiliation{Max-Planck-Institut f\"ur Kernphysik, Saupfercheckweg~1, 
69117 Heidelberg, Germany}

\date{\today}

\begin{abstract}
Electric dipole-forbidden transitions of nuclei interacting with super-intense laser fields are investigated considering stable isotopes with suitable low-lying first excited states. Different classes of transitions are identified, and all magnetic sublevels corresponding to the near-resonantly driven nuclear transition are included in the description of the nuclear quantum system. We find that large transition matrix elements and convenient resonance energies qualify  nuclear $M1$ transitions as good candidates for the coherent driving of nuclei. We discuss the implications of resonant interaction of intense laser fields with nuclei beyond the dipole approximation for the controlled preparation of excited nuclear states and important aspects of possible experiments aimed at observing these effects. 

\end{abstract}

\pacs{25.20.Dc, 29.30.Kv, 42.50.-p, 42.55.Vc}

\keywords{nuclear quantum optics, x-ray laser, electric dipole-forbidden transitions}

\maketitle


\section{Introduction}

The direct interaction of laser fields with nuclei has been considered impossible for a long time, most studies being focused on indirect reactions in which parts of the electronic shell or plasmas act as secondary particles~\cite{Pretzler, Ledingham, Izumi}.   The  nucleus-laser interaction matrix elements are small~\cite{Matinyan}, the polarization of nuclei is ineffective, as the populations of the hyperfine levels are nearly equal even at very low temperatures, and the energies of the photons are typically several orders of magnitude away from those of the nuclear transitions. 
Meanwhile, synchrotron radiation evolved into a versatile source of light which enables one to carry out a great variety of nuclear spectroscopy experiments (e.g. Ref.~\cite{Sturhahn,Hyp_volume} and references therein). Crystal monochromators even allow the generation of narrow-bandwidth partially coherent light out of synchrotron radiation, such that, e.g., quantum beats could be observed in time spectra of nuclear forward scattering~\cite{vanB1}.

It is still generally believed that higher power, brilliance and temporal and transverse coherence at low wavelengths, possibly combined with ultrashort pulses, will open unprecedented perspectives for related experiments in nuclear physics. The most promising candidates for such a fourth-generation light source are free electron lasers (FEL) such as the European X-Ray Free-Electron Laser (XFEL)~\cite{TESLADesign,tesla}, the Linac Coherent Light Source (LCLS) at SLAC~\cite{LCLSDesign}, the SPring-8 Compact SASE Source (SCSS) in Japan~\cite{SCSSDesign}, the BESSY high-gain harmonic generation FEL in Berlin~\cite{bessy} or the fourth generation light source 4GLS in Daresbury~\cite{4gls} (an extensive list with currently operating FELs and facilities under development can be found in Ref.~\cite{fels}).    
Recently, in view of the forthcoming novel light sources, the direct resonant laser-nucleus interaction has been investigated \cite{Thomas1}, considering super-intense laser fields both in the visible and in the x-ray frequency regime. The authors of Ref.~\cite{Thomas1} have shown that present and upcoming high-frequency laser facilities, especially when combined with a moderate acceleration of the target nuclei to match photon and transition frequency, do allow for resonant or off-resonant \cite{Thomas2} laser-nucleus interaction. 
This opens the possibility of optical measurements of nuclear properties such as the transition frequency and the nuclear  transition matrix element, further developing the field of nuclear quantum optics. In particular, the coherence of the laser light expected from new sources such as the European XFEL \cite{TESLADesign,tesla}, with photon energies envisaged up to 12.4~keV, is the essential feature which may allow to attain nuclear coherence or interference phenomena reminiscent of atomic quantum optics, such as nuclear Rabi oscillations, photon echoes or more advanced quantum optical schemes  \cite{Scully} in nuclei.

While in atomic systems typically electric dipole transitions
dominate the light-matter interaction because of both their
relevant transition frequencies and their larger dipole moments, 
in the case of nuclei, this issue is less obvious. Due to the symmetrical structure of nuclei, multipolarities of higher order than electric dipole ($E1$) are the rule, and often electronic transitions of the atom, such as internal conversion (IC), compete strongly with radiative transitions of excited nuclei \cite{Baldwin}.
The spectrum of $E1$ transitions is limited to few low-lying nuclear
excited states with small reduced transition probabilities,
and giant resonances, at energies of several MeV, which are not 
directly accessible nowadays with the laser and are problematic even when considering the acceleration of the target nucleus.  In analogy to atomic quantum optics, until now only laser-driven electric dipole transitions in nuclei have been studied \cite{Thomas1,Thomas2}, completely disregarding the large choice of electric dipole-forbidden transitions.

We therefore investigate in this paper electric dipole-forbidden transitions of nuclei interacting with super-intense laser fields and show that, unlike in atomic quantum optics,  these are suitable candidates for resonant coherent driving of nuclei. Since in most cases near-resonant intermediate states are not available and the present laser intensities suppress non-resonant multiphoton transitions, our study is devoted to one-photon transitions. We consider laser-driven transitions between the ground  and first excited state of suitable stable or long-lived isotopes, including all the magnetic sublevels in a multilevel model of the nuclear system interacting with the laser field. 
In order to account for dipole-forbidden transitions, the plane waves describing the electromagnetic field are expanded in spherical waves characterized by well-defined multipolarity and parity. The nuclear interaction matrix element is expressed with the help of the reduced transition probabilities, for which we use experimental values. 
The interaction matrix elements and population inversion for electric and magnetic dipole and electric quadrupole transitions are compared for several stable or long-lived nuclei with low-lying first excited states. This comparison includes both cases in which the direct laser-nucleus interaction is by itself possible as well as  nuclear systems for which moderate acceleration of the target nuclei is required to match nuclear transition and photon energies.
We find that $M1$ transitions are prospective candidates for resonant coherent driving of nuclei due to their large transition probabilities and suitable energies. A comparison between electric dipole-allowed and forbidden transitions shows that, unlike in atomic cases,  for similar transition energies the interaction matrix elements are often of the same order of magnitude. This result considerably expands the choice of suitable isotopes since electric dipole transitions from the ground state in the keV region are very rare.

This paper is organized as follows: in Section II, we derive the laser-nucleus interaction Hamiltonian used in this study and calculate the interaction matrix element for higher multipoles, focussing on $E2$ and $M1$ transitions. Section III presents the numerical results for the interaction matrix elements and population inversion of the nuclear multilevel system and discusses aspects of a possible experimental observation of the laser driving of nuclei. We conclude with a short summary.


\section{\label{theory} Theoretical formalism}
 For applications in quantum optics, the density matrix theory has been utilized successfully 
to describe the dynamics of the atomic system interaction with laser fields including relaxation~\cite{Scully}. In the following, we apply the density matrix formalism to the few-level nuclear system driven by the electromagnetic field. We use a semiclassical description in which the nucleus is treated as a quantum  few-level system and the field is treated classically. The equations of motion for the density matrix are derived in Section \ref{Optical_Bloch}. The laser-nucleus interaction  Hamiltonian is detailed in Section \ref{me}, where the electromagnetic field is expanded in spherical waves taking into account the specific terms responsible for the considered dipole-forbidden transitions. 


\subsection{\label{Optical_Bloch} Basic quantum dynamics}

The interaction Hamiltonian between the nucleus and the electromagnetic field can be written in the semiclassical approximation as
\begin{equation}
H_{I}=-\frac{1}{c}\int d^3r \:  \vec{j}(\vec{r})\cdot \vec{A}(\vec{r},t)\, ,
\label{H_int}
\end{equation}
where $\vec{j}(\vec{r})$ represents the nuclear charge current, $c$ is the speed of light, and the vector potential of the field in the Coulomb gauge is given by
\begin{equation}
\vec{A}(\vec{r},t)=\frac{c}{\omega_k}E_k e^{-i\omega_k t} e^{-i\vec{k}\cdot\vec{r}} {\vec{e}\,}^*_{\vec{k}\sigma}+{\rm c.c.}
\label{a}
\end{equation}
The electric field amplitude is denoted by $E_k$, and the photons are characterized by the wave vector $\vec{k}$, the frequency $\omega_k$, and the polarization $\sigma$. Furthermore, $ \vec{e}_{\vec{k}\sigma}$ is the polarization vector. As an example we consider circularly polarized light, so that $\sigma=\pm 1$. The case of linearly polarized light can be treated in a similar manner.  

We restrict the Hilbert space of the nucleus to the excited state $|e\rangle$ and the ground state $|g\rangle$ characterized by the angular momenta $I_e$ and $I_g$, respectively, including their magnetic sublevels $M_e$ and $M_g$. 
The nuclei assumed to be initially 
in the state $|g\rangle$ are irradiated with the intensity $\mathcal{I}(t)$ starting from time $t=0$. 
The dynamics of the density matrix $\rho$ is determined by
the Master equation~\cite{Scully} 
\begin{equation}
i\hbar \frac{\partial}{\partial t} \rho=[H_0+H_I,\rho]+\mathcal{L}\rho\, ,
\label{eqmot}
\end{equation}
where $H_0$ denotes the unperturbed nuclear Hamiltonian operator
and the Lindblad operator $\mathcal{L}$ describes the various spontaneous relaxation channels.

The density matrix element corresponding to the states $i,j\in\{e,g\}$, that are characterized by the magnetic quantum numbers $M_i$ and $M_j$, respectively, is denoted by $\rho_{ij}$. 
The optical Bloch equations evaluate to 
\begin{align}
\label{bloch}
\frac{\partial}{\partial t}\rho_{gg}(M_g)=& -\frac{2}{\hbar}{\mathrm{Im}}\left(\sum_{M_e}\rho_{ge}(M_g,M_e)e^{i\omega_kt}\langle I_e, M_e|H_I|I_g, M_g\rangle\right)
\nonumber \\
&+\sum_{M_e}\gamma_s(M_g,M_e)\rho_{ee}(M_e)\, ,
\nonumber \allowdisplaybreaks[2] \\
\frac{\partial}{\partial t}\rho_{ge}(M_g,M_e)=&-i\Delta\rho_{ge}(M_g,M_e)+\frac{i}{\hbar}\left (\rho_{gg}(M_g)-\rho_{ee}(M_e) \right )e^{-i\omega_kt}
\langle I_g,M_g|H_I|I_e,M_e \rangle
\nonumber \\
&-\frac{\gamma_s(M_g,M_e)}{2}\rho_{ge}(M_g,M_e)-\gamma_{dec}\rho_{ge}(M_g,M_e)\, ,
\nonumber \allowdisplaybreaks[2]\\
\frac{\partial}{\partial t}\rho_{ee}(M_e)=&\frac{2}{\hbar}{\mathrm{Im}}\left(\sum_{M_g}\rho_{ge}(M_g,M_e)e^{i\omega_kt}\langle I_e,M_e|H_I|I_g,M_g \rangle\right)
\nonumber \\
&-\rho_{ee}(M_e)\sum_{M_g}\gamma_s(M_g,M_e)\,,
\end{align}
where the rapidly oscillating terms in the off-diagonal elements have been eliminated by moving to a rotating frame. The spontaneous decay of the excited level  $|e\rangle$   depends on the magnetic sublevels of the nuclear states. With $\Gamma_s$ being the total decay rate of the $|e\rangle$ state, summed up over all the possible values of $M_e$ and $M_g$, the partial decay rate  $\gamma_s(M_g,M_e)$ is given by the Clebsch-Gordan coefficients $C(j_1\ j_2\ j_3;m_1\ m_2\ m_3)$ via
\begin{equation}
\gamma_s(M_g,M_e)=\frac{2I_e+1}{2L+1}[C(I_g\ I_e\ L;M_g\ -\!M_e \ M)]^2\:\Gamma_s\, ,
\end{equation}
where $L$ and $M$ denote the photon multipolarity and total angular momentum projection, respectively. 
In the optical Bloch equations (\ref{bloch}), $\gamma_{dec}$ stands for an additional dephasing rate to model laser field pulses with limited coherence times, while $\Delta=\omega_0-\omega_k$ is the detuning of the laser frequency with respect to the nuclear transition frequency $\omega_0$. The system of differential equations with the given initial conditions can be solved once the matrix element of the interaction Hamiltonian is known. In the following section we calculate the interaction matrix element, which is also contained in the expression of the Rabi frequency of the system.

\subsection{Interaction matrix element}
\label{me}
In atomic physics, the electromagnetic field is usually described by the plane-wave expansion, such that the vector potential has the expression given in Eq.~(\ref{a}). The photons are characterized by the wave number $k$, the  propagation direction $\hat{k}$ and the polarization $\sigma$. In the case of nuclear transitions, however, it is more convenient to describe the photons in terms of multipolarity and parity. We therefore  expand the plane wave in multipoles to account for photons characterized by the angular momentum $L$, its projection $M$ and the parity $\lambda$. We follow the formalism described in \cite{Rose} and  express the product in the expression of the interaction Hamiltonian in Eq.~(\ref{H_int}) with the help of the electric $\mathcal{E}$ and magnetic $\mathcal{M}$ multipole fields,
\begin{equation}
\vec{j}\cdot{\vec{e}\,}^*_{\vec{k}\sigma} e^{-i\vec{k}\cdot\vec{r}} =\sum_{LM} \sqrt{2\pi(2L+1)}(-i)^L D^L_{M\,-\sigma}(\hat{k})\vec{j}(\vec{r})\cdot\left( \vec{A}_{LM}^{\mathcal{M}}(\vec{r})+i\sigma\vec{A}_{LM}^{\mathcal{E}}(\vec{r})\right)\, ,
\label{multipole}
\end{equation} 
where $D^L_{M\,-\sigma}(\hat{k})$ is the rotation matrix (characterized by the photon angular momentum projection $M$ and polarization $\sigma$) associated to the rotation of the coordinate system which transforms  the direction of the $z$-axis to $\hat{k}$.  The magnetic $\vec{A}_{LM}^{\mathcal{M}}(\vec{r})$ and electric $\vec{A}_{LM}^{\mathcal{E}}(\vec{r})$ multipole fields are given by
\begin{eqnarray}
\vec{A}_{LM}^{\mathcal{M}}(\vec{r})&=&j_L(kr)\vec{Y}_{LL}^M(\theta,\varphi)\, ,
\nonumber \\
\vec{A}_{LM}^{\mathcal{E}}(\vec{r})&=&\sqrt{\frac{L+1}{2L+1}}j_{L-1}(kr)\vec{Y}_{LL-1}^M(\theta,\varphi)-\sqrt{\frac{L}{2L+1}}j_{L+1}(kr)\vec{Y}_{LL+1}^M(\theta,\varphi)
\nonumber \\
&=& -\frac{i}{k}\nabla\times \left(j_L(kr)\vec{Y}_{LL}^M(\theta,\varphi)\right)\, ,
\end{eqnarray}
where $j_L(kr)$ are the spherical Bessel functions and  $\vec{Y}_{LL}^M(\theta,\varphi)$ denote the vector spherical harmonics.
Typically, only a few terms in the sum in Eq.~(\ref{multipole}), i.e., one or two multipole orders, are needed for the accurate calculation of the interaction matrix element. In the case of magnetic multipole transitions, the interaction Hamiltonian has the form
\begin{equation}
H_I=-\frac{E_k}{\omega_k}e^{-i\omega_kt} \sqrt{2\pi}\sum_{LM}(-i)^L\sqrt{2L+1}D^L_{M\,-\sigma}(\hat{k})\int d^3r j_L(kr)\vec{j}(\vec{r})\cdot\vec{Y}^M_{LL}(	\theta,\varphi)\, .
\end{equation}
We consider in the following that the direction of the incoming photons is parallel to that of the $z$-axis, so that $D^L_{M\,-\sigma}(\hat{k})=\delta_{M,-\sigma}$, where $\delta_{i,j}$ is the Kronecker delta symbol. As the wavelength of the radiation is large compared to the nuclear radius ($kR_0\ll 1$), we use the long wavelength approximation, keeping only the first order term $(kr)^L$ in an expansion in $(kr)$ in the spherical Bessel function.
The interaction matrix element then becomes
\begin{eqnarray}
\langle I_e, M_e|H_I|I_g, M_g\rangle&=&E_ke^{-i\omega_kt}\sqrt{2\pi}\sum_{L}(-i)^{L-1}\sqrt{\frac{(2L+1)(L+1)}{L}}
\nonumber \\
&\times&
\frac{k^{L-1}}{(2L+1)!!}\langle I_e, M_e|M_{L\,-\sigma}|I_g, M_g\rangle\, ,
\label{intme}
\end{eqnarray}
where $M_{L\,-\sigma}$ is the nuclear magnetic multipole moment operator corresponding to the total angular momentum $L$ and its projection $-\sigma$, defined as \cite{Ring}
\begin{equation}
M_{LM}=\frac{1}{c(L+1)}
\int d^3r
(\vec{r}\times\vec{j}(\vec{r}))\cdot\vec{\nabla}(r^LY_{LM}(\theta,\varphi))\, .
\end{equation}
The symbol $!!$ in Eq.~(\ref{intme}) denotes the double factorial given by
$n!! = n(n-2)(n-4)\cdots \kappa_n$, where $\kappa_n$ is 1 for odd $n$
and $2$ for even $n$. 
Since the multipole moments  are spherical tensor operators, we can factor out the magnetic quantum number dependence of their matrix elements with the help of the  Wigner-Eckart theorem \cite{Edmonds}, 
\begin{equation}
\langle I_e, M_e|M_{L\,-\sigma}|I_g, M_g\rangle=\frac{(-1)^{I_g-M_g}}{\sqrt{2L+1}}C(I_e\ I_g\ L;M_e\ -M_g\ -\sigma)\langle I_e\|M_{L}\|I_g\rangle\, .
\end{equation}
The reduced matrix element of the magnetic multipole moment $\langle I_e\|M_{L}\|I_g\rangle$ is now independent of the angular momentum substructure and can be related to the reduced transition probability $B$ by \cite{Ring}
\begin{equation}
B({\mathcal{M}} L, I_g\to I_e)=\frac{1}{2I_g+1}|\langle I_e\|M_{L}\|I_g\rangle|^2
\, .
\end{equation}
The interaction matrix element for a certain multipolarity $L$ can then be written as 
\begin{eqnarray}
\langle I_e, M_e|H_I|I_g, M_g\rangle&\sim&E_ke^{-i\omega_kt}\sqrt{2\pi}\sqrt{\frac{(L+1)}{L}}\frac{k^{L-1}}{(2L+1)!!}C(I_e\ I_g\ L;M_e\ -M_g\ -\sigma) 
\nonumber \\
&\times& \sqrt{2I_g+1}\sqrt{B({\mathcal{M}}L,I_g\to I_e) }\, ,
\end{eqnarray}
omitting an overall phase factor which is irrelevant, since the population dynamics of the few-level system depends on the squared modulus of the interaction matrix element.
In a similar manner, by identifying the nuclear electric multipole operator in the expression of the interaction Hamiltonian for electric transitions, we obtain for the matrix element the expression
\begin{eqnarray}
\langle I_e, M_e|H_I|I_g, M_g\rangle&\sim&E_ke^{-i\omega_kt}\sqrt{2\pi}\sqrt{\frac{(2L+1)(L+1)}{L}}\frac{k^{L-1}}{(2L+1)!!}C(I_e\ I_g\ L;M_e\ -M_g\ -\sigma)
\nonumber \\
&\times& \sqrt{2I_g+1}\sqrt{B({\mathcal{E}}L,I_g\to I_e) }\, .
\end{eqnarray}


\section{\label{results} Results and discussion}

The interaction matrix elements and population inversion for electric and magnetic dipole and electric quadrupole transitions are calculated and compared for several stable or long-lived nuclei with low-lying first excited states. This comparison includes both cases in which the direct laser-nucleus interaction is by itself possible (the nuclear transition energy $E$  allows for  direct resonant interaction, $E<12.4$~keV), as well as  nuclear systems for which moderate acceleration of the target nuclei is required to match nuclear transition and photon energies ($E>12.4$~keV). Signal rates 
are calculated envisaging realistic experimental parameters of the laser and nuclear targets. An overview of the important  light source parameters is given in Section \ref{light_sources}, while the numerical results are present and discussed in Section \ref{num-results}.

\subsection{Light sources}
\label{light_sources}
Powerful high-frequency coherent light sources are the key element for achieving direct laser-nucleus interactions. In the x-ray energy range, FEL is currently the only suitable lasing me\-cha\-nism, exploiting, for example, the self-amplified spontaneous emission (SASE) of free electrons accelerated and propagated through an undulator. Since the electrons in the FEL are not bound to atoms and thus not limited to specific transitions, the wavelength of the FEL is tunable over a wide range depending on accelerator energy and undulator parameters. At the future European XFEL,  photon energies  up to 12.4~keV are envisaged, with an average spectral brightness~\cite{Nomenclature} of 1.6$\times 10^{25}$ photons/(sec$\cdot$mrad$^2\cdot$mm$^2\cdot$0.1$\%$bandwidth) in pulses of 100~fs at a repetition rate of 40~kHz \cite{tesla,TESLADesign}. The SPring-8 Compact SASE Source (SCSS) in Japan aims at a slightly lower spectral brightness, with similar photon energies~\cite{SCSSDesign}. The Linac Coherent Light Source (LCLS) at SLAC \cite{LCLSDesign} will deliver photons with energies up to 8.2~keV with an average spectral brightness of 2.7$\times 10^{22}$ photons/(sec$\cdot$mrad$^2\cdot$mm$^2\cdot$0.1$\%$bandwidth).  
Other FEL light sources such as the BESSY high-gain harmonic generation FEL in Berlin \cite{bessy} or the 4th generation light source 4GLS in Daresbury \cite{4gls} can reach only a lower photon energy, up to 1~keV or 100~eV, respectively.  As the number of laser-driven nuclear transitions per pulse is small, the pulse repetition rate is a crucial value for the magnitude of the nuclear excitation effect to be observed. The European XFEL has the largest repetition rate of 40 kHz, while the SCSS at Spring-8 and the LCLS FEL at SLAC have only 60 Hz and 120 Hz, respectively. The light sources with lower photon energy at BESSY and 4GLS can provide pulses with a repetition rate of approximately 1kHz. 

Since the European XFEL is designed to reach the highest photon energy, pulse repetition rate and average brilliance, we use  its parameters to numerically estimate the possible magnitude of direct laser-nucleus interaction. The peak intensity of the laser can be obtained from the peak power divided by the area of the focal spot, $ \mathcal{I}=6\times 10^{15}$ W/cm$^2$.
 For a  pulse of $10^{12}$ photons with an energy of 12.4~keV, the total energy spread $\Gamma_L$ of about 10~eV is given by the spectral bandwidth of 0.08~$\%$. The nuclear transition width is many orders of magnitude smaller than the photon energy spread, implying that only a fraction of the incoming photons will actually fulfill the resonance condition. From the flux of photons resonant within the transition width of the excited state $\Gamma_N$, we deduce the effective laser intensity $\mathcal{I}_{\rm eff}$,
\begin{equation}
\mathcal{I}_{\rm eff}=\frac{\Gamma_N}{\Gamma_L}\mathcal{I}\, .
\end{equation}
As far as the magnitude of the peak intensity is concerned, we distinguish two cases. 
In the first case, we consider an optimization of the number of signal photons scattered off of the nuclei for a resonant laser field. Since the intensity of the considered light sources typically does not allow for full Rabi flopping in the laser-nuclei interaction due to the small width of the nuclear resonances, high average intensities rather than peak intensities are desirable. For laser sources strong enough to achieve an inversion of the nuclear population (via a so-called $\pi$-pulse~\cite{Scully}), the intensity should be adjusted such that the population evolution does not exceed the complete inversion, as otherwise the signal photon rate is reduced. Second, if quantum optical scenarios which require a Rabi frequency comparable to the radiative decay rates are considered, high peak intensities are mandatory. The feasibility of quantum optical schemes with present laser sources will be discussed in more detail in the following section.

If the target nuclei are accelerated, the boosted electric field $E_N$ and the laser frequency $\nu_N$ are given by
\begin{eqnarray}
E_N&=&(1+\beta)\gamma E_L\, ,
\nonumber \\
\nu_N&=&(1+\beta)\gamma \nu_L\, ,
\end{eqnarray}
where the subscripts $N$ and $L$ stand for the rest frame of the nucleus and the laboratory frame, respectively. The reduced velocity $\beta$ and the Lorenz factor $\gamma$ are chosen such that the laser and nuclear transition frequencies match, e.g. a nuclear transition of 100~keV with 12.4~keV laser photons correspond to a Lorentz factor of $\gamma=4.1$. In the boosted reference frame, the effective laser intensity is multiplied by the square of the boost factor, 
\begin{equation}
\label{ieff}
\mathcal{I}_{\rm eff}^N=(1+\beta)^2\gamma^2 \mathcal{I}_{\rm eff}^L\, .
\end{equation}

For the particular case of very low-lying first nuclear excited states in the eV regime, such as the case of the isomeric state of $^{229}_{90}\mathrm{Th}$ which will be discussed later in this section, conventional light sources as well as FEL lasers can be used. Among the conventional UV lasers, Vulcan \cite{vulcan} at the Rutherford Appleton Laboratory  and  PHELIX \cite{phelix1,phelix2} at the GSI in Darmstadt can reach intensities of up to  $\mathcal{I}=10^{21}$~W/cm$^2$.  The drawback of these light sources is  the low repetition rate of order of one pulse per 10~min. 
A better repetition rate of the order of MHz can be achieved, e.g.  by the VUV-FEL at the 4GLS in Daresbury \cite{4gls} 
 with a photon energy of up to 10~eV,  and an average brightness of  5$\times 10^{21}$ photons/(sec$\cdot$mrad$^2\cdot$mm$^2\cdot$0.1$\%$bandwidth).

\subsection{\label{num-results}Numerical results}
We calculate the interaction matrix elements for a number of nuclear systems involving $E2$ and $M1$ transitions between the ground and first excited nuclear states. Particular interest is devoted to nuclei in which the energy of the first  excited state allows for a direct interaction with the laser without requiring the acceleration of the target nucleus for the given European XFEL parameters. Namely, we investigate the cases of the $^{83}_{36}\mathrm{Kr}$, $^{137}_{57}\mathrm{La}$, $^{151}_{62}\mathrm{Sm}$, $^{169}_{69}\mathrm{Tm}$ and $^{187}_{76}\mathrm{Os}$ isotopes, which have  first excited states lying below   12.4~keV. These isotopes have stable or long-lived ground states, so that a reduction of the full level structure to the nuclear ground state and the first excited state in a few-level approximation is well justified.
Each of these states is split into its respective magnetic sublevels. For  comparison, a low-lying $E1$ transition  at 6.238 keV in $^{181}_{73}\mathrm{Ta}$ is also investigated. For the moment, in the framework of a general discussion, we assume  that the spontaneous decay of the excited nuclear state occurs only radiatively. This corresponds to nuclei in bare ions or cases in which nuclear de-excitation via IC is not possible. 
The radiative decay rate $\Gamma_s$ can be calculated using the reduced transition probabilities \cite{Ring},
\begin{equation}
\Gamma_s(\lambda,L)=\frac{8\pi(L+1)}{L[(2L+1)!!]^2}\left(\frac{E}{\hbar c}\right)^{2L+1}B(\lambda L,I_e\to I_g)\, ,
\end{equation}
where $\lambda$ stands for electric ${\mathcal{E}}$ or magnetic ${\mathcal{M}}$, and $E$ denotes the transition energy.
Values for the reduced transition probabilities, the interaction matrix elements and the radiative decay rates are given in Table \ref{table_1}. The reduced interaction matrix element denotes the quantity
\begin{equation}
\langle I_e \|H_I\|I_g \rangle=E_{\rm eff}\sqrt{2\pi}\sqrt{\frac{(L+1)}{L}}\frac{k^{L-1}}{(2L+1)!!}\sqrt{2I_g+1}\sqrt{B( {\mathcal{M}}L,I_g\to I_e) }
\end{equation}
for the case of magnetic transitions and 
\begin{equation}
\langle I_e \|H_I\|I_g \rangle=E_{\rm eff}\sqrt{2\pi}\sqrt{\frac{(2L+1)(L+1)}{L}}\frac{k^{L-1}}{(2L+1)!!}\sqrt{2I_g+1}\sqrt{B({\mathcal{E}}L,I_g\to I_e) }
\end{equation}
for electric transitions. Here, $E_{\rm eff}$ is the effective electric field amplitude, corresponding to the effective laser intensity $\mathcal{I}_{\rm eff}$. For these cases, the XFEL is assumed to deliver photons in resonance with the nuclear transition. The energy levels and the experimental values for $B(\lambda L,I_e\to I_g)$ were taken from \cite{Kr83,La137,Sm151,Tm169,Ta181,Os187}. The reduced transition probabilities for the emission and the absorption of a gamma ray, respectively, are related through the formula
\begin{equation}
B(\lambda L,I_g\to I_e)=\frac{2I_e+1}{2I_g+1}B(\lambda L,I_e\to I_g)\, .
\end{equation}
In the cases of the  $^{83}_{36}\mathrm{Kr}$, $^{151}_{62}\mathrm{Sm}$, $^{169}_{69}\mathrm{Tm}$ and $^{187}_{76}\mathrm{Os}$ isotopes, we neglect the weak multipole mixing of the nuclear transitions, considering only the dominant $M1$ component. The multipole mixing ratio values given in Ref.~\cite{toi} support this approximation, which affects with less than one percent the accuracy of the calculated signal rate.

\begin{table*}[tb]
\caption{\label{table_1} Reduced interaction matrix elements and spontaneous decay rates for a choice of isotopes with first excited states lying below 12.4~keV. In the fifth and sixth columns, $E$ and   $B(\lambda L,I_g\to I_e)$ denote the nuclear excitation energy and the reduced transition probability, respectively, taken from Refs.~\cite{Kr83,La137,Sm151,Tm169,Ta181,Os187}. }
\begin{ruledtabular}
\begin{tabular}{cccccccc}

Isotope  & $I_g$& $I_e$ & $L$ & $E$ (keV) & $B(\lambda L,I_g\to I_e)\ (\mathrm{e}^2\mathrm{fm}^{2L})$  &  $\langle I_e \|H_I \|I_g \rangle$ (eV) & $\Gamma_s$ (1/s) \\
\hline
$^{83}_{36}\mathrm{Kr}$ & $9/2^+$ & $7/2^-$ & M1+E2 &  9.396 & $1.49\times 10^{-4}$ &$1.61\times 10^{-10}$ &$2.51\times 10^5$  \\
$^{137}_{57}\mathrm{La}$ & $7/2^+$ & $5/2^+$ & M1 &  10.56 & $2.51\times 10^{-5}$ &$2.50\times 10^{-11}$ &$6.37\times 10^4$ \\
$^{151}_{62}\mathrm{Sm}$ & $5/2^-$ & $3/2^-$ & M1+E2 &  4.821 &$7.74\times 10^{-5}$  & $7.07\times 10^{-11}$&$2.10\times 10^4$ \\
$^{169}_{69}\mathrm{Tm}$ & $1/2^+$ & $3/2^+$ & M1+E2 &  8.410 &$1.24\times 10^{-3}$   & $3.79\times 10^{-10}$ & $5.96\times 10^5$\\
$^{187}_{76}{\rm Os}$& $1/2^-$ & $3/2^-$ & M1(+E2) & 9.746  &$1.26\times 10^{-3}$  & $3.84\times 10^{-10}$ & $9.43\times 10^5$\\
\hline
$^{181}_{73}{\rm Ta}$& $7/2^+$ & $9/2^-$ & E1 & 6.238  &$5.18\times 10^{-6}$  & $6.84\times 10^{-12}$ & $1.59\times 10^3$\\

\end{tabular}
\end{ruledtabular}
\end{table*}

The values for the reduced interaction matrix element presented in Table \ref{table_1} show that for low-lying nuclear levels with energies of approximately 10 keV, $M1$ transitions are well-suited for direct interaction with strong laser fields. The largest interaction matrix element occurs in the case of the 9.746 keV  $M1$ transition of $^{187}_{76}\mathrm{Os}$. 

\begin{table*}[tb]
\caption{\label{table_2} Reduced interaction matrix elements and spontaneous decay rates for a choice of isotopes with low-lying first excited states above 12.4~keV. In the fifth and sixth columns, $E$ and   $B(\lambda L,I_g\to I_e)$ denote the nuclear excitation energy and the reduced transition probability, respectively, taken from Refs.~\cite{Eu153,Dy161,Raman,Yb173,Eu151}.}
\begin{ruledtabular}
\begin{tabular}{cccccccc}

Isotope  & $I_g$& $I_e$ & $L$ & $E$ (keV)  & $B(\lambda L,I_g\to I_e)\ (\mathrm{e}^2\mathrm{fm}^{2L})$  &  $\langle I_e \|H_I \|I_g \rangle$ (eV) & $\Gamma_s$ (1/s) \\ 
\hline
$^{153}_{62}{\rm Sm}$& $3/2^+$ & $3/2^-$ & E1 &  35.843 & $>3.50\times 10^{-2}$ &$2.27\times 10^{-7}$  & $2.55\times 10^9$ \\
$^{153}_{63}{\rm Eu}$& $5/2^+$ & $5/2^-$ & E1 &  97.429 & $1.80\times 10^{-3}$  & $1.06\times 10^{-7}$ & $2.64\times 10^9$\\
$^{161}_{66}{\rm Dy}$& $5/2^+$ & $5/2^-$ & E1 &  25.651 & $2.65\times 10^{-4}$  &$1.08\times 10^{-9}$  & $7.09\times 10^6$ \\
\hline
$^{156}_{64}{\rm Gd}$& $0^+$ & $2^+$ & E2 & 88.966& $4.64\times 10^{4}$  & $3.27\times 10^{-9}$ & $6.32\times 10^7$
 \\
$^{162}_{66}{\rm Dy}$& $0^+$ & $2^+$ & E2 & 80.660 & $5.35\times 10^{4}$  &  $2.54\times 10^{-9}$ & $4.46\times 10^7$\\
$^{238}_{92}{\rm U}$& $0^+$ & $2^+$ & E2 & 44.910 & $12.09\times 10^{4}$  &  $6.72\times 10^{-10}$ & $5.50\times 10^6$\\
\hline
$^{173}_{70}{\rm Yb}$& $5/2^-$ & $7/2^-$ & M1 & 78.647  & $3.07\times 10^{-3}$  & $5.94\times 10^{-8}$ & $1.81\times 10^9$\\
$^{151}_{63}{\rm Eu}$& $5/2^+$ & $7/2^+$ & M1 &  21.532 & $2.17\times 10^{-4}$  & $3.14\times 10^{-10}$ & $2.62\times 10^6$\\
$^{165}_{67}{\rm Ho}$& $7/2^-$ & $9/2^-$ & M1 & 94.700  & $6.76\times 10^{-3}$  & $2.26\times 10^{-7}$ & $7.42\times 10^9$\\
\hline
\end{tabular}
\end{ruledtabular}
\end{table*}

In order to extend the comparison between electric dipole allowed and forbidden transitions we also consider typical examples of stable or long-lived isotopes with the first excited state at energies covering the range from 12.4 keV up to 100 keV. 
Since the width of the first excited state determines the effective intensity of the laser field, we chose for several energy ranges and transition types  isotopes with short-lived excited levels. The effective intensity of the laser $\mathcal{I}_{\rm eff}$ is calculated according to Eq.~(\ref{ieff}), assuming that the XFEL delivers 12.4~keV photons that are boosted in the nuclear reference frame to reach the transition frequency.
 In Table \ref{table_2} we present reduced interaction matrix elements and spontaneous decay rates for the
$E1$ transitions of $^{153}_{62}{\rm Sm}$, $^{153}_{63}\mathrm{Eu}$ and  $^{161}_{66}\mathrm{Dy}$, the $0^+ \! \to\! 2^+$ E2 transitions of $^{156}_{64}\mathrm{Gd}$, $^{162}_{66}\mathrm{Dy}$ and $^{238}_{92}\mathrm{U}$ and finally the $M1$ transitions of $^{173}_{70}\mathrm{Yb}$, $^{165}_{67}\mathrm{Ho}$ and $^{151}_{63}\mathrm{Eu}$. All these isotopes have stable ground states, except for $^{153}_{62}{\rm Sm}$ which has a mean lifetime of $\tau=46.27$ h \cite{toi}, and the very long-lived  $^{238}_{92}\mathrm{U}$ ground state with mean-life of approximately $4\times 10^9$~yr. The lifetime of the nuclear ground state is  important when considering possible experimental observation of the coherent driving of nuclei, because it affects the sample preparation. While the interaction matrix element is the largest for the $E1$ transition in 
$^{153}_{62}{\rm Sm}$, this isotope has the disadvantage of its unstable ground state. On the other hand, the $M1$ transition in the stable $^{165}_{67}{\rm Ho}$ isotope can be a good candidate for the experimental observation of nuclear quantum optics effects. The higher order term in the wave number $k$ causes the $E2$ transitions to have smaller interaction matrix elements, as shown in Table \ref{table_2}. 

\begin{figure}
\begin{center}
\includegraphics[width=0.60\textwidth]{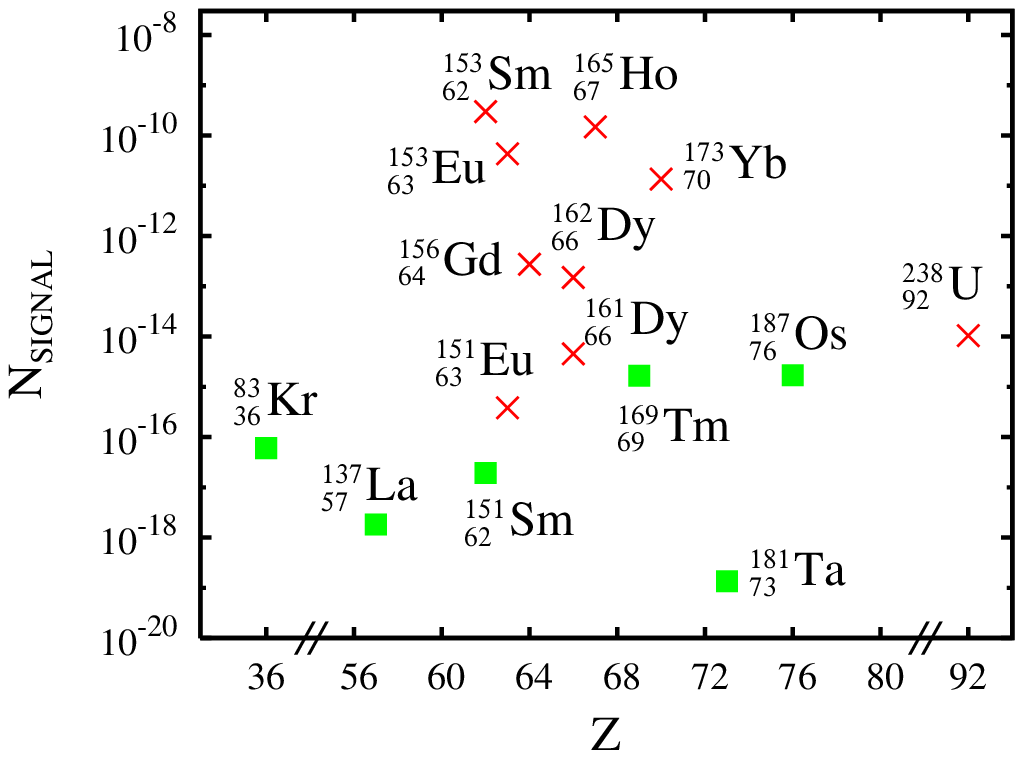}
\caption{\label{points} (Color online) Number of signal photons per nucleus per laser pulse $N_{\rm SIGNAL}$ for several isotopes with first excited states below 12.4~keV (green squares) and above  12.4~keV (red crosses). The results are plotted versus the atomic number $Z$. The European XFEL considered has a pulse duration of 100~fs and an average brilliance of 1.6$\times 10^{25}$ photons/(sec$\cdot$mrad$^2\cdot$mm$^2\cdot$0.1$\%$bandwidth).
        }
\end{center}
\end{figure} 

By solving the optical Bloch equations in Eq.~(\ref{bloch}) we can obtain the population of the excited state after one pulse, considering as initial condition that the ground state population is equally distributed between the corresponding magnetic substates \cite{OlgaPRL}. In the scenario of laser interacting with bare nuclei, after the laser pulse,  the excited nuclei decay  back to the ground state with the emission of a signal photon. Thus, the sum $\rho_{ee}$ of the excited state population $\rho_{ee}(M_e)$ over the possible magnetic sublevels $M_e$ is equal to the (unpolarized) signal photon rate per nucleus per pulse $N_{\rm SIGNAL}$ for the driven nuclear transition. We present in Figure \ref{points} the signal photon rate for the isotopes considered in Tables \ref{table_1} and \ref{table_2} for nuclear transitions with energies above and below 12.4~keV using the European XFEL laser parameters. The highest signal rate is $N_{\rm SIGNAL}=3\times 10^{-10}$ photons per nucleus per pulse for the case of the  $E1$ transition of the $^{153}_{62}{\rm Sm}$ isotope, two times larger than the one for the $M1$ transition of $^{165}_{67}{\rm Ho}$. 

\begin{figure}
\begin{center}
\includegraphics[height=5.5cm]{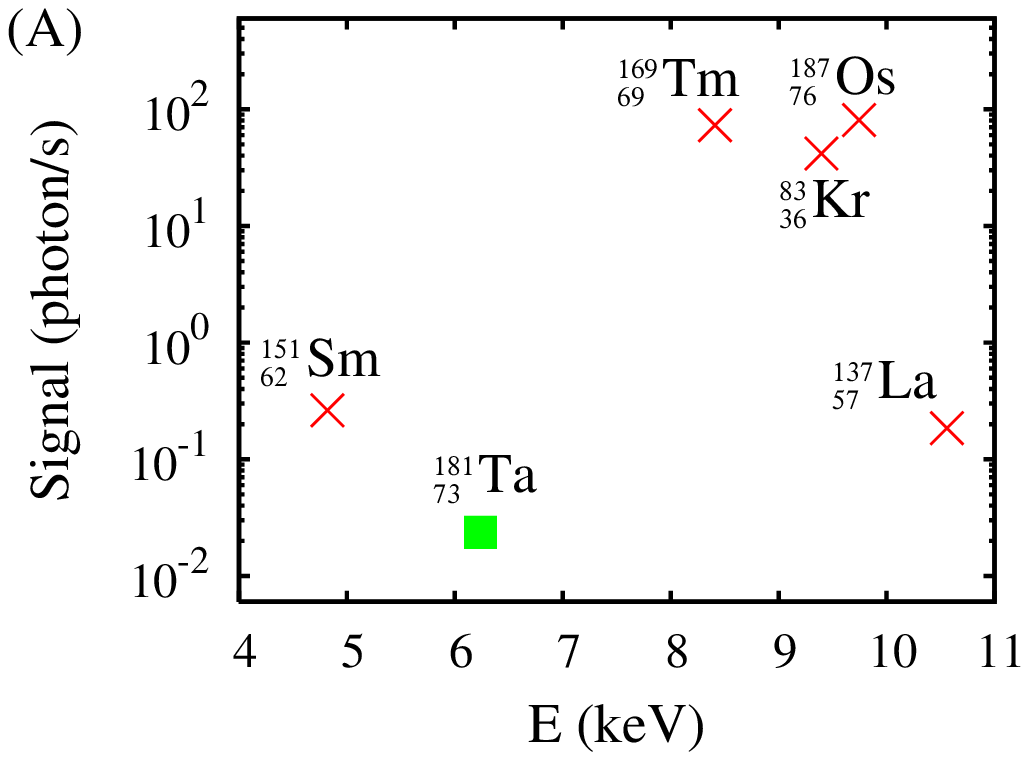}
\hspace*{0.3cm}
\includegraphics[height=5.5cm]{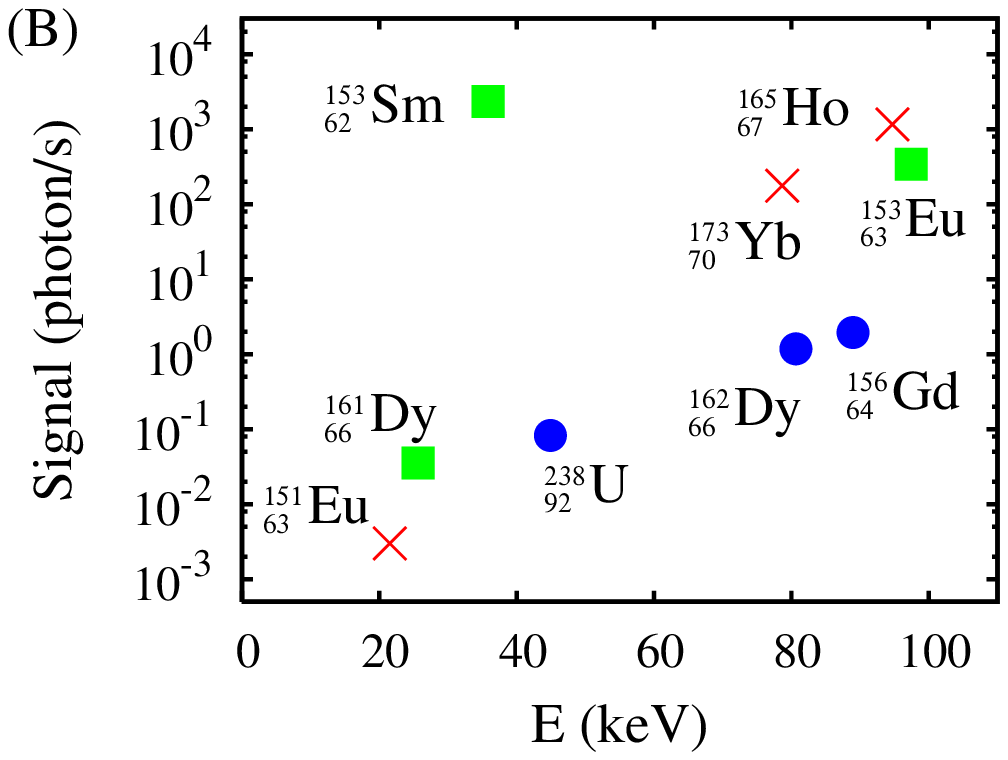}
\caption{\label{signal} (Color online) Signal photons per second for laser driving of nuclear transitions with energies (A) below 12.4~keV and (B) above 12.4~keV. The energy of the nuclear transition is given on the abscissa. The green squares denote $E1$ transitions, the blue circles $E2$ transitions and the red crosses  $M1$ transitions.
The considered experimental laser and target parameters are discussed in Sec.~\ref{num-results}.}
\end{center}
\end{figure} 

Generally, due to the very narrow widths of excited nuclear levels lying below 12.4~keV, the number of photons in the laser pulse that fulfill the resonance condition for the considered isotopes is small. The signal photon rates are for these cases up to ten orders of magnitude smaller than the ones for transitions above 12.4~keV. With regard to a possible experiment aimed at the observation of the direct laser-nucleus interaction, isotopes with low-lying first excited states below 12.4~keV have nevertheless the advantage that the laser can directly provide photons at the resonance energy. Solid targets which have a very high density of nuclei can therefore be used for such an experiment. In this case, however, the presence of atomic electrons opens the IC decay channel for the nuclear excited state. The branching ratio for the nuclear decay can be related to the IC coefficient $\alpha$, defined as the ratio between the IC and radiative decay rates. Due to the small energies of the nuclear excited states, the IC decay is more probable, with $\alpha$ values between 17.09 for the  9.396~keV transition of $^{83}_{36}\mathrm{Kr}$ and 920 for the 4.821~keV transition of $^{151}_{62}\mathrm{Sm}$~\cite{Kr83,La137,Sm151,Tm169,Ta181,Os187}. The photon signal per nucleus per pulse is then given by the fraction of the excited state population that will decay radiatively, $\rho_{ee}/(1+\alpha)$. In order to estimate the signal photon rate per second, we assume a focal diameter of the laser of 20~$\mu$m together with a repetition rate of 40~kHz \cite{TESLADesign,tesla} and a density of $10^{20}$ nuclei/cm$^2$ in the solid state target. 
In Figure~\ref{signal}(A) we present the photon signal per second for the $M1$ and $E1$ nuclear transitions presented in Table~\ref{table_1}. The largest signal rate of 80.75~photons/second  corresponds to the 9.746~keV $M1$ transition of $^{187}_{76}{\rm Os}$.

 It should be noted, however, that the presence of the electrons in the solid target complicates the identification of laser driving of nuclei. The photon scattering off of the  crystal electrons creates a considerable background.  Furthermore, the irradiation of a strong XFEL laser on a solid target would lead to the formation of a plasma, invoking many other nuclear excitation or decay mechanisms involving high-energy charged particles, such as  nuclear excitation by electron capture \cite{Goldanskii,Palffy,Palffy2} or  nuclear excitation by electron transition \cite{Tkalya}. Nuclear excitation and decay mechanisms in plasmas have been  the subject of several studies  \cite{Tkalya_pl,Morel,Harston} which show the importance of taking into account the electron-nucleus coupling. Given the narrow widths of the low-lying nuclear excited states, solid state effects like inhomogeneous broadening of the nuclear line \cite{Balko} may play an important role. We would like to stress, however, that similar difficulties have been experimentally overcome in the case of M\"ossbauer spectroscopy using synchrotron radiation \cite{vanB1}, which led to the observation of interesting effects such as nuclear superradiance \cite{vanB2,vanB3} and nuclear exciton echos \cite{vanB4}.

The driving of nuclear levels with energies larger than 12.4~keV requires acceleration of the target nuclei in order to match the photon and nuclear transition energy. This restricts the target to ion beams, which have a small density in comparison with solid targets. With $2.5\times 10^{10}$ particles in a bunch length of $\tau=50$~ns in an ion beam of 2~mm diameter as target \cite{Tahir}, the particle density yields $5.3\times 10^8$~cm$^{-3}$. For the new Synchrotron SIS100 that  will be built in the future at FAIR \cite{fair} the beam parameters yield a particle density of $10^{11}$ ions/cm$^3$ \cite{Tahir}. Using this value to estimate the number of signal photons per second we obtain $2.35\times 10^3$~s$^{-1}$ for the $E1$ transition in $^{153}_{62}{\rm Sm}$ and $1.16\times 10^3$~s$^{-1}$ for the $M1$ transition in $^{165}_{67}{\rm Ho}$. Here, the calculation of the photon rate per second assumes the matching of the ion and laser pulse repetition rates.
The signal photon rates per second for the $E1$, $E2$ and $M1$ transitions given in Table~\ref{table_2} are presented in Figure~\ref{signal}(B). In addition to the larger signal photon rates, the experimental background is also more advantageous for such transitions, since fewer competing effects are present. In the case of bare ions, for instance,  the IC nuclear decay channel  can be suppressed completely.

 For the case of laser driving of the transition to the first excited state of $^{165}_{67}{\rm Ho}$, the boosted XFEL intensity of approximately $\mathcal{I}=4\times 10^{17}$~W/cm$^2$ corresponds to an effective intensity  $\mathcal{I}_{\rm eff}=10^{11}$~W/cm$^2$.  As a result of laser driving of the nuclear transition, signal photons corresponding to the radiative decay of the nuclei can be observed. The dynamics of the nuclear few-level system strongly depends on the laser intensity.
 For the considered XFEL laser intensity, the nuclear system remains on average almost in the ground state (the excitation rate per nucleus per laser pulse is  $\rho_{ee}=1.5\times 10^{-10}$ for $^{165}_{67}{\rm Ho}$), such that only few nuclei are excited in each laser pulse. With increasing intensity the nuclear excited state population starts to oscillate rapidly at about $\mathcal{I}=10^{28}$~W/cm$^2$. A $\pi$ pulse that would directly transfer all nuclei to the excited state with no further oscillations can be found around $\mathcal{I}_{\pi}=10^{27}$~W/cm$^2$.  In analogy to atomic quantum optics, at $\mathcal{I}=10^{39}$~W/cm$^2$, when the Rabi frequency equals the transition frequency,  strong-field effects beyond the rotating wave 
 approximation are expected to occur. However, note that with driving fields approaching the Schwinger intensity of about $10^{29}$~W/cm$^2$, other processes such as pair creation and vacuum screening  become relevant and may prevent observation of the nuclear excitation.

Finally, we would like to address as an interesting case study the possible excitation of the isomeric state of $^{229}_{90}{\rm Th}$ that lies only 7.6~eV above the ground state \cite{Beiersdorfer}. The metastable state is assumed to decay to the ground state by an $M1$ transition with a lifetime of approximately 5~h. While this lifetime corresponds to a very narrow width of the nuclear excited state, one may expect that in solid targets homogeneous line broadening can be of the order of $10^{-12}$~eV \cite{Jos,Jos_Cous}. Unfortunately, since the decay of the isomeric state has not been observed directly \cite{Irwin,Richardson,Utter,Shaw}, the reduced transition probability is unknown, thus making it impossible to give a realistic estimation of the laser-nucleus interaction matrix element. Detection of the laser-induced transition can be achieved by probing the hyperfine structure of a transition in the electronic shell, as it has been proposed in  Ref.~\cite{Peik}.  Since the energy value of the $^{229\rm m}{\rm Th}$ isomer was recently corrected from 3.5~eV to 7.6~eV, numerous studies considering laser-assisted or laser-driven coupling between the nucleus and the electronic shell (see, for instance, Refs.~\cite{Typel,Tkalya_x,Karpeshin,Kalman}) need revision.


\section{\label{summary} Summary}

We have considered the direct interaction between strong laser fields and nuclei focusing in particular on the driving of electric dipole-forbidden transitions. 
The main motivation for investigating such transitions is the limited variety of stable nuclear systems with $E1$ transitions in the interesting low-energy range. While $E1$ transitions with first excited states lying below 100~keV exist, the systems for which the ground state is stable or long-lived are limited to only a few cases.  We have investigated laser-driven $M1$ and $E2$ transitions for a number of stable systems with energies above and below the maximum photon energy of 12.4~keV envisaged at the European XFEL, and have compared the results with corresponding $E1$ cases. We found that on average, laser-driving of $M1$ transitions is as effective as the driving of $E1$ transitions. In particular for excited levels lying below 12.4~keV,  $M1$ transitions are the most promising for laser excitation. This is in contrast to atomic quantum optics, where higher multipole transitions are strongly suppressed. This equivalence of different multipole transitions in direct nucleus-laser interactions considerably enhances the range of suitable isotopes. Our analysis further shows that generally, in the search for a suitable experimental candidate system, the strength of the nuclear transition, usually given as the reduced transition probability, is a key quantity in estimating the magnitude of the effect. Finally, we discussed two possible scenarios for the experimental observation of direct laser-nucleus interaction. For transition energies below 12.4~keV, no target acceleration is required, such that high-density solid targets are preferable. We found signal rates of up to 80 photons/second
for the considered isotopes with low transition frequencies. Plasma effects occurring in solid state targets are, however, likely to complicate the experimental observation and require an adequate theoretical description. 
 Above the maximum source photon energy, the use of accelerated ion beams is
required. These have a lower target density than solid-state
targets, but also a much reduced background. Using realistic parameters, we found for this case signal photon rates of up to $10^4$ photons per second.


\begin{acknowledgments}

The authors would like to thank Thomas J. B\"urvenich for fruitful discussions.

\end{acknowledgments}

\bibliography{palffy_resub}

\end{document}